\documentclass[global,twocolumn]{svjour}

\usepackage{amsmath,amssymb}
\usepackage{graphicx}
\usepackage{color}
\usepackage[english]{babel}

\begin{document}

\title{Solid-state laser system for laser cooling of Sodium}

\author{Emmanuel Mimoun\inst{1} \and Luigi de Sarlo\inst{1} \and Jean-Jacques Zondy\inst{2} \and Jean Dalibard\inst{1} \and Fabrice Gerbier\inst{1}}

\institute{Laboratoire Kastler Brossel, ENS, UPMC, CNRS, 24 rue
Lhomond, 75005 Paris, France.\and INM-CNAM, 61 rue du Landy, 93210 La Plaine Saint Denis,
France.}

\maketitle
\begin{abstract}
We demonstrate a frequency-stabilized, all-solid laser source at 589~nm with up to 800 mW
output power. The laser relies on sum-frequency generation from two laser sources at
1064~nm and 1319~nm through a PPKTP crystal in a doubly-resonant cavity. We obtain conversion efficiency as high as 2~W/W$^2$ after careful optimization of the cavity parameters. The output wavelength is tunable over 60 GHz, which is sufficient to lock on the Sodium D$_2$ line. The robustness, beam quality, spectral narrowness and tunability of our source make it an alternative to dye lasers for atomic physics experiments with Sodium atoms.  \keywords{37.10.De -- 42.65.Ky}
\end{abstract}

%

\section{Introduction}

Among the many atomic species that can be brought to quantum
degeneracy, Sodium benefits from low inelastic losses and a
relatively large elastic cross-section, allowing the production of
large ($> 10^8$ atoms) Bose-Einstein condensates~\cite{davis95a,hau1998a,naik2005a,magalhaes2005a,streed2006a,dumke2006a,stam2007a}, or
degenerate Fermi clouds by thermalization with the Sodium gas~\cite{hadzibabic2003a}. A current drawback of using Sodium is the
necessity of using dye lasers to reach the resonant wavelength of
589.158~nm (Sodium D$_2$ transition). Although technically
well-mastered, dye laser systems are expensive, hardly transportable
and comparatively difficult to maintain and operate, justifying the need for alternatives as solid-state lasers.

 In addition, new laser sources in the yellow spectral region find applications
outside the domain of laser cooling. In fact, the generation of
Sodium resonant radiation has been mainly driven by the astronomy
community, with the development of high-power 589~nm lasers to
create artificial ``beacon'' stars by exciting the mesospheric
sodium layer~\cite{jeys89a,moosmueller97a,vance98a,bienfang2003a,janousek2005a,feng2004a,mildren2004a,georgiev2006a,sinha2006a}.
Other possible applications for lasers in the yellow spectral region
include Laser-induced detection in the atmospheric range (LIDAR)~\cite{fugate1991a}, eye surgery or dermatology~\cite{janousek2005a}.

In the literature, several methods for generating continuous-wave
(cw) laser light around 589~nm have been reported, including
sum-frequency mixing of two infrared lasers around 1319~nm and 1064~nm~\cite{jeys89a,moosmueller97a,vance98a,bienfang2003a,janousek2005a,nishikawa2009a},
frequency-doubling of a Raman fiber laser~\cite{feng2004a,mildren2004a,georgiev2006a}, or sum-frequency mixing
of two fiber lasers around 938~nm and 1535~nm~\cite{sinha2006a}. Applications to laser cooling typically require powers of several hundred mW
to 1~W, the possibility to tune the laser to the Sodium resonance,
and a linewidth much smaller than the $\Gamma=2\pi\times 9.8~$MHz
natural linewidth of the cooling line.

In a recent paper, we have reported on the realization of a laser source
suitable for laser cooling of Sodium~\cite{mimoun2008a}. In the present article, we present an exhaustive account of our experimental approach. Our laser source is based on sum
frequency generation (SFG) from 1064~nm and 1319~nm lasers. SFG is a
second order non-linear optical process, in which two pump beams
with frequencies $\omega_1$ ($\lambda_1=1064~$nm) and $\omega_2$ ($\lambda_2=1319~$nm) produce a signal beam
with frequency $\omega_3 =\omega_1 + \omega_2$. We implement this sum frequency technique using commercial, solid-state infrared lasers.
The 1064~nm and 1319~nm sources are monolithic solid-state lasers built upon an
Yttrium Aluminium Garnet (YAG) Non-Planar Ring Oscillator (NPRO)
crystal. We also tested another configuration in which the 1064~nm laser is replaced by an external cavity laser diode boosted by a single-mode fiber amplifier. This led to poorer performances attributed to misbehaviour of the amplifier, and this configuration was not pursued further in our work. The non-linear medium used is a periodically poled
potassium titanyl phosphate crystal (pp-KTiOPO$_4$ or PPKTP), with a poling
period chosen to achieve first-order quasi-phase matching (QPM) near room
temperature~\cite{myers1995a,boyd2003a}. In single-pass
configuration, the conversion efficiency is still too small to reach
the output power required for laser cooling. To circumvent this
problem, the crystal is enclosed in a doubly-resonant build-up
cavity to enhance the conversion efficiency. Doing so, we reach an overall
power conversion efficiency $\alpha_{\rm cav}\approx2~$W/W$^2$, where the
conversion efficiency is defined through $P_3=\alpha_{\rm } P_1 P_2$, with $P_{1,2,3}$ the power at each wavelength. This is to be compared to the value for single-pass conversion efficiency, $\alpha_{\rm sp}\approx0.022~$W/W$^2$. In terms of photon fluxes, about 92\% of the photons of the weakest source which enter the cavity are converted~\cite{mimoun2008a}.

The article is organized as follows. Section~\ref{sec:expsetup} gives an overview of our experimental setup. Section~\ref{sec:spmeas} recalls the
main features of the process of sum frequency generation, and
presents our results in a simple single-pass configuration. Section~\ref{sec:cavity_theory} discusses the experimental realization of a doubly resonant cavity and its optimization to achieve near-unit conversion efficiency. Section~\ref{sec:yellow} characterizes the main properties of the laser source obtained at
589~nm.

\section{Experimental setup}
\label{sec:expsetup}

\begin{figure*}[ht!]
\centering
\includegraphics{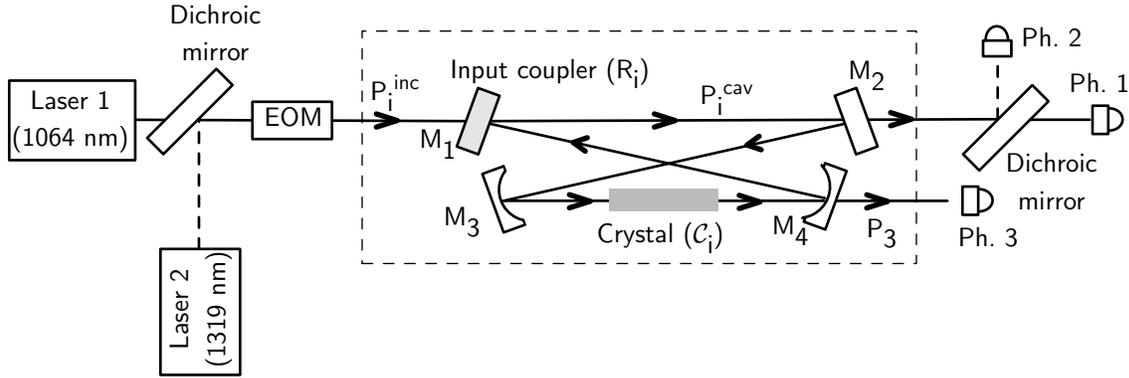}
\caption{Overview of the laser system. Mirrors M1 (with reflectance R$_i$ at wavelength $\lambda_i$) and M2 are flat,
while mirrors M3 and M4 are concave with a radius of curvature
of 10 cm to allow focusing in the crystal, in which a fraction ${\cal C}_i$ of the power at wavelength $\lambda_i$ is converted. The incoming power at wavelength $\lambda_i$ is noted $P_i^{\rm inc}$, and the intra-cavity power $P_i^{\rm cav}$. Ph. stands for Photodiode and EOM for Electro-Optic phase Modulator.}
\label{fig:Cavity}
\end{figure*}

Our experimental setup is represented in Fig.~\ref{fig:Cavity}. The pump laser sources at $1064~$nm and $1319~$nm are non-planar ring oscillator YAG lasers (Innolight GmbH, Germany) with an instantaneous spectral linewidth narrower
than $10~$kHz and output power of  $1.1~$W and $500~$mW, respectively. Both lasers are collimated to a $1/e^2$ radius around $1~$mm and combined on a dichroic mirror. After passing through an electro-optical phase modulator (EOM) operating near $f_{\rm mod}\approx1~$MHz, the beams  are focused to match their spatial profiles with the fundamental spatial mode of the resonant cavity. 

The cavity is built in a bow-tie planar configuration, with highly reflecting mirrors $M_{2}, M_{3}, M_{4}$  and an input coupler $M_{1}$ with lower reflectivities chosen such as to optimize the intracavity conversion (see section \ref{sec:cavity_theory} below). Mirrors $M_{1},M_{2}$ are plane, while $M_{3}, M_{4}$ are concave with radius of curvature $R_{c}=100~$mm. At the crystal location, the $1064~$nm ($1319~$nm) beam is focused to a waist $w_{1}=45~\mu$m ($w_{2}=47~\mu$m). This corresponds to almost equal Rayleigh lengths in the crystal $z_{R,i}=\pi\ n_i\ w_i^2/\lambda_i\simeq 10$~mm, with the refractive indices $n_1\simeq1.83$ and $n_2\simeq1.82$ for PPKTP.  For our configuration, this choice offers a good trade-off between increasing the nonlinear conversion efficiency and avoiding detrimental effects such as thermal lensing. The cavity geometry is chosen to avoid transverse mode degeneracies, allowing to excite the fundamental Gaussian mode only and suppress higher order modes. 

The PPKTP crystal used for SF mixing was manufactured at the Royal Institute of
Technology of Stockholm (KTH). Its length is  $L=20~$mm, with a
poling period $\Lambda=12.36~\mu$m. The use of a periodically poled
crystal allows to reach QPM conditions only with temperature tuning
(see section~\ref{sec:theory1}). The crystal is mounted in a copper
case with a Peltier thermo-electric cooler element. The case temperature is controlled by a
standard Proportional-Integral-Derivative regulator with
better than 10~mK stability. Using the phase-matching curve
calculated in section~\ref{sec:theory1}, we estimate that this corresponds to output power drifts
less than 1\%.

\section{Single-pass measurements}
\label{sec:spmeas}
In this section, we discuss first our measurements in a single-pass configuration, {\it i.e.} without enhancement cavity. As shown later, this measurement is critical to optimize the resonant cavity parameters to reach maximal conversion efficiency. We first recall for completeness the theoretical results relevant to such measurements, first in the simple case where the pumps are plane waves and then in the more realistic situation where they are Gaussian beams. We then discuss our measurements - from which we derive a nonlinear coefficient d$_{33}\simeq16$pm/V for PPKTP. 

\subsection{Plane wave model}
\label{sec:theory1}

In this section, we recall the basic features of SFG using a simplified
theoretical framework~\cite{boyd2003a}. The starting point to describe the propagation in the nonlinear
crystal are Helmholtz equations, including nonlinear
polarization terms. We introduce the complex amplitudes $a_i$,
related to the electric field strengths by $$E_i=\sqrt{\frac{2 Z_0
\hbar\omega_i}{n_i}} f_i({\bf \rho}) a_i(z) e^{i(k_i z-\omega_i t)}$$
and to the powers by $P_{i}=\hbar \omega_{i} |a_{i}|^2$. Here $k_i$
is the momentum of a photon with frequency $\omega_i$ in a medium
with index of refraction $n_i$, $z$ is the direction of propagation of
light, ${\bf \rho}$ is the transverse coordinate, $f_i$ denotes the
area-normalized transverse mode for each beam, and
$Z_0=\sqrt{\mu_0/\epsilon_0}$ is the impedance of vacuum.

As a first approximation, we neglect the spatial profile of the
laser beams and set $f_i({\bf \rho}) = S^{-1/2}$, with a cross-section
$S$ identical for all beams. The Helmholtz equation for the harmonic
wave $a_3$ then reads~\cite{boyd2003a}
\begin{eqnarray}
\label{eqn:a3}
\frac{d a_3}{dz} & = & - i\gamma a_1 a_2 e^{-i\ \Delta k\ z},
\end{eqnarray}
where the non-linear coupling coefficient $\gamma$ can be written as
\begin{eqnarray}
\label{eqn:gamma}
    \gamma & = & \left(\frac{2 \hbar \omega_1 \omega_2 \omega_3 Z_0^3
    \epsilon_0^2 d^2}{S n_1 n_2 n_3}\right)^{1/2}.
\end{eqnarray}
Here $d$ denotes the nonlinear coefficient which characterizes
the efficiency of the nonlinear process~\footnote{We assume here that
the laser polarizations are parallel and aligned with the principal
axis of the non-linear medium characterized by the largest
non-linear coefficient $d_{33}$.}, and $\Delta k=k_3-k_1-k_2$ is the phase
mismatch parameter. In a bulk crystal, $d$ can be treated as
constant over the crystal length (neglecting possible defects and
impurities). In contrast, a periodically poled crystal is
characterized by an alternating permanent ferromagnetization~\cite{myers1995a}. As a result, $d$ is a periodic function of the
position $z$ in the poling direction, with spatial period $\Lambda$.
As such, it can be expanded as a Fourier series, $d(z)=d_{33}\sum_n c_n
e^{i q_{n} z}$, where $q_n=2\pi n/ \Lambda$. Significant conversion
only takes place when the QPM condition
$q_n=\Delta k$ occurs for some integer $n$. Here we only consider the first term of the series, $n=1$ (first order QPM). For a 50\% poling duty-cycle, the Fourier coefficient $c_{1}$ is $2 /\pi$ and the nonlinear coefficient $d$ in Eq.(\ref{eqn:gamma}) becomes an effective coefficient $d_{\rm pp}=2/\pi\ d_{33}$. Hence, the maximum efficiency is lower than for a bulk crystal with
perfect phase matching by a factor $4/\pi^2\approx0.4$.

Assuming low conversion, we solve for $a_3$ in the undepleted pumps
approximation, $a_{i}(z)\approx a_{i}(0)$, for $i=1,2$. For a crystal of length $L$, this gives the generated power at 589~nm as
\begin{eqnarray}\label{P3qpm}
P_3  & = & \alpha_{\rm sp} P_1 P_2 {\rm sinc}^2\left(\frac{\left(\Delta
k-\frac{2\pi}{\Lambda}\right) L}{2}\right),
\end{eqnarray}
where ${\rm sinc}(x)=\sin (x)/x$ and where the maximal single-pass conversion efficiency $\alpha_{\rm sp}$ is
\begin{equation}  \label{alphaqpm}
\alpha_{\rm sp} = \frac{\gamma^2 L^2}{\hbar}\frac{\omega_3}{\omega_1
\omega_2}.
\end{equation}
As we will see, the single-pass efficiency $\alpha_{\rm sp}$ is the critical parameter to allow optimization of the resonant cavity. 
The argument of the sinc function in Eq.(\ref{P3qpm}) depends on temperature through the various refractive indices (the case of KTP has been studied experimentally in~\cite{fradkin1999a,emanueli2003a}). Therefore, by adjusting the temperature one can reach the quasi-phase matching condition $\Delta k = 2\pi /\Lambda$ which maximizes the conversion efficiency. In practice, the period $\Lambda$ is chosen so that this condition is fulfilled near room temperature. 

\subsection{Boyd-Kleinmann theory}

Instead of collimated beams, experiments use focused Gaussian beams in order to reach
high intensities, and hence efficient conversion. Non-linear
processes with Gaussian waves were studied in a seminal paper by Boyd and Kleinmann~\cite{boyd1968a}, where explicit expressions were given for
the conversion efficiency in the undepleted pumps approximation (see also~\cite{zondy1997a}). The general expressions are rather complex, but they can
be drastically simplified by assuming identical Rayleigh
lengths $z_{\rm R}$ for the three beams. Indeed, both infrared beams are resonant in the cavity, which implies that their confocal parameter is the same, essentially determined by the geometry of the cavity. Although the output beam generated by SFG is not resonant, it is generated only in the regions where both pump beams overlap significantly so that approximating its spatial mode by a Gaussian beam with the same confocal parameter as the infrared ones is a reasonable assumption~\cite{boyd2003a}. 

With Gaussian beams, the coefficient $\gamma$ defined in
Eq.(\ref{eqn:gamma}) becomes a function of $z$ proportional to the overlap integral $I(z) =\int d^{(2)}{\bf \rho} f_1 f_2 f_3^\ast$ between the different modes $f_i$. For Gaussian waves with waists  $w_i$ at the crystal center, and Rayleigh length $z_{R}=\pi\ n_i\ w_i^2/\lambda_i$, this can be calculated explicitly. After some
rearrangement, the expression for the output power can be writtten as $P_3=\alpha_{\rm sp} P_1 P_2$, where the
single-pass conversion efficiency $\alpha_{\rm sp}$ reads
\begin{eqnarray}
\alpha_{\rm sp} & = & Z_1 \frac{d_{\rm pp}^2 L}{\lambda_3^3}h\left(a,b,c\right).
\end{eqnarray}
Here
\begin{eqnarray} Z_1=\frac{32 \pi Z_0 }{
\lambda_1 \lambda_2
\left(\frac{n_1}{\lambda_1}+\frac{n_2}{\lambda_2}+\frac{n_3}{\lambda_3}\right)^2}\approx
2.15~{\rm k\Omega}
\end{eqnarray}
has the dimension of an impedance and the dimensionless
function $h$
\begin{eqnarray}\label{f}
h\left(a,b,c\right)=\frac{1}{4a}\left| \int_{-a}^a \frac{e^{i b
\tau}}{(1+i\tau)(1+i c \tau)}d\tau \right|^2
\end{eqnarray}
is the so-called Boyd-Kleinman factor. The latter depends on the reduced variables $a=\frac{L}{2z_R}$, $b=\left(\Delta k-\frac{2\pi}{\Lambda}\right)z_R$, and $c=\Delta k w_{\rm eff}^2/z_R$, with $w_{\rm eff}^{-2}=(\pi/z_R)\sum_{i}n_i/\lambda_i$. We can further write $c  =\frac{\Delta k\ w_{\rm eff}^2}{z_R}= \left( b + \frac{1}{a}\frac{\pi L}{\Lambda}\right) \times \frac{w_{\rm
eff}^2}{z_R L} \times a$, showing that the function $h$ depends only on the variables
$a$ and $b$ once the wavelengths, crystal length and crystal period
are fixed~\footnote{In principle, $Z_0$ and $c$ depend weakly on temperature as $\Delta
k$ through the dependance of the indices. One finds that over a
temperature range of $20^\circ-100^\circ$ the relative variations do
not exceed a few $10^{-4}$. Thus, we can safely consider $Z_0$ and
$c$ as constants for the rest of the calculations.}. Since $
\frac{w_{\rm eff}^2}{z_R L}\sim
\frac{\lambda_3}{2\pi n_3 L} \ll 1$, $c=0$ can be assumed, and the integral $h\left(a,b,c\right)$ is well approximated by 
\begin{eqnarray}\label{fast}
h\left(a,b,0\right)=\frac{1}{4a}\left|
\int_{-a}^a \frac{e^{i b \tau}}{1+i\tau}d\tau \right|^2.
\end{eqnarray}
There are two limiting cases of interest :
\begin{enumerate}
\item {\bf Collimated beams, $z_R \gg L$ or $a \ll 1$:} in this case
we find the sinc function familiar from the plane wave case~(see Eq.(\ref{P3qpm})),
\begin{eqnarray}
h\left(a,b,0\right)\approx a\ {\rm sinc}^2\Bigl((b+1)a\Bigr).
\end{eqnarray}
\item {\bf Focused beams, $z_R \ll L$ or $a \gg 1$:}
for tightly focused beams, the length $L$ of the crystal naturally
drops out of the problem. One finds that $h$ tends to a limit function
\begin{displaymath}
h\left(a\gg 1,b,0\right) \approx \left\{\begin{array}{ll}
\frac{\pi^2}{a} e^{-2b},& ~~b>0,\\
0,& ~~b<0.
\end{array}\right.
\end{displaymath}
\end{enumerate}
The experimental procedure of changing the temperature (which changes $\Delta k$) corresponds to searching for the maximum  $h^\ast (a)$ of $h(a,b,0)$ as a function of $b$ for a fixed $a$~\cite{boyd1968a}. The
optimum phase mismatch is offset from the plane wave result ($\Delta
k = 2\pi /\Lambda$) by a quantity on the order of $z_R^{-1}$, a consequence of the Gouy phase accumulated as the beams pass
through a focus in the crystal. The optimal focusing corresponds to the maximum of $h^\ast$, which is found for $a^\ast=L/2z_{R}\approx 2.84$ ( $h^\ast(2.84)\approx 1.06$). This optimum is quite loose, as $h^\ast >1$ for $1.5\lesssim a \lesssim 5$. 


\subsection{Results for single pass operation}

\begin{figure}[htbp]
    \centering
        \includegraphics[width=8cm]{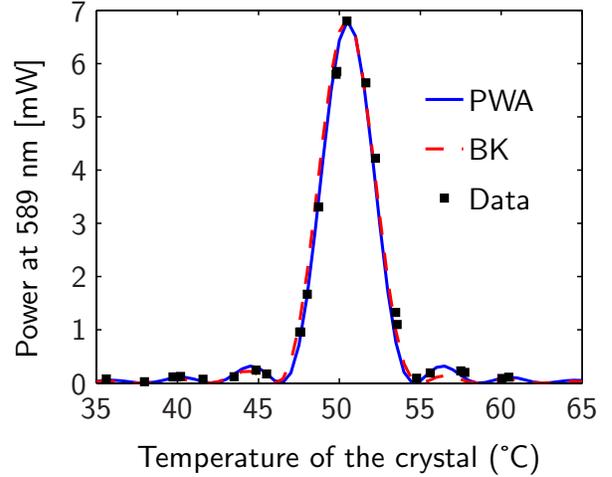}
    \caption{Power produced at 589~nm in a single-pass configuration when the temperature of the PPKTP crystal is varied. The plot is fitted for a value of the poling period $\Lambda=12.32~\mu$m, for the Plane Wave Approximation (solid line, PWA) and the Boyd Kleinmann theory (dashed line, BK). The peak value measured for the power produced at the output of the cavity is 7~mW. Considering the losses due the optics used to separate yellow light from infrared, we infer a total produced power of 8.5~mW.}
    \label{fig:Fits_temperature}
\end{figure}

Our experimental configuration corresponds to a configuration where $z_{R}\approx10~$mm, or $a\approx2$ for a $20~$mm long crystal. For this parameter, the shape of the function $h$ is very close to the sinc function predicted by the plane wave model (see Fig.~\ref{fig:Fits_temperature}). To perform single-pass measurements, we use the same setup as in Fig.~\ref{fig:Cavity} but remove the mirror $M_{4}$ from the cavity. We find that the optimum temperature $T_{\rm QPM}^{(\rm{
mes})}\approx 50^{\circ}$C for our crystal samples. For this temperature, the refractive indices are $n_1\approx1.83$, $n_2\approx1.82$, and
$n_3\approx1.87$. Using BK theory to fit the data, we measure an efficiency
$\alpha_{\rm sp}^{\rm mes}\approx 0.022$~W$^{-1}$ ($P_{3}=9$~mW), obtained
for normal incidence on the crystal using $P_{2}=440$~mW and $P_{1}
=940~$mW~\footnote{This is the highest value obtained so far and differs from the measurements presented in Fig.~\ref{fig:Fits_temperature}, which we have taken with another laser source at $1064~$nm of poorer quality, and another crystal.}. The variation of the output power with the power of both
infrared lasers are found to be linear, confirming the validity of
the undepleted pumps approximation for a single pass operation.
Boyd-Kleinmann theory predicts a value $\alpha_{\rm sp}^{\rm BK}\approx 0.021~$W/W$^2$ for $d_{33}\approx 16~$pm/V, which is in a good agreement with values for the non-linear coefficient found in the literature~\cite{arie1997a,popov2000a,targat2005a}. Note also that this is quite close to the optimal value $\alpha^\ast\approx0.023~$W/W$^2$ which would be obtained for slightly tighter focusing. Applying the equations (\ref{P3qpm}-\ref{alphaqpm}) obtained in the plane wave approximation, one would expect $\alpha\approx 0.022$, taking for the cross section $S$ the average of the waist of a gaussian beam over the length L of the crystal : $S=\frac{1}{L}\int^{L/2}_{-L/2}\pi w^2(z)/2\ dz$, with  $w(0)=45$~$\mu$m. This highlights the usefulness of these theories for quantitative predictions.

The PPKTP crystals used in our experiments  was ordered with a
poling period $\Lambda=12.36~\mu$m. Using the
values given in Refs.~\cite{fradkin1999a,emanueli2003a} for
the temperature and wavelength dependence of the refractive indices, we calculate a QPM temperature $T_{\rm
QPM}^{(\rm {calc})}\approx 28^{\circ}$C, apparently far from the measured $T_{\rm QPM}^{(\rm{
mes})}\approx 50^{\circ}$C (see
Fig.~\ref{fig:Fits_temperature}). We note however that the quasi-phase-matching temperature is rather
sensitive to the exact value of the period. Using the same
wavelength and temperature dependence for the refraction indices, we
find that the measured $T_{\rm QPM}^{(\rm{ mes})}$ corresponds to a
spatial period $\Lambda=12.32~\mu$m.

\section{Intra-cavity conversion}
\label{sec:cavity_theory}

\subsection{Definition of the optimization problem}

After having characterized the single-pass sum-frequency process, we turn to the
cavity setup. The presence of the cavity enhances the infrared
lasers intensities at the crystal location. For the geometry shown in
Fig.~\ref{fig:Cavity}, the intracavity power at resonance can be written for
each infrared laser as~\cite{siegman1986a}
\begin{equation}\label{IcavNL}
P_i^{\rm cav}=\frac{T_{i}}{\left(1-\sqrt{R_{i} \cdot (1-\delta_{i})\cdot
C_i}\right)^2} P_i^{\rm
inc},\quad i=1,2.
\end{equation}
In Eq.(\ref{IcavNL}), $P_i^{\rm cav}$ denotes the circulating intracavity power, $P_i^{\rm
inc}$ the incident power coupled into the fundamental mode of the cavity, $R_{i}, T_{i}$ denote the input coupler (mirror $M_{1}$) reflectance and transmittance ($R_{i}+T_{i}+L_{i}=1$, with $L_{i}$ a loss coefficient), $\delta_{i}$ denote the passive losses after one round trip, excluding the input coupler (i.e. finite reflectances of the other mirrors, and losses in the crystal), and $C_i$ accounts for the nonlinear conversion. To
evaluate $C_i$, we use conservation laws
for the photon fluxes which state that $|a_1|^2+|a_3|^2$ and
$|a_2|^2+|a_3|^2$ are constant along the crystal length (in the
absence of absorption). This corresponds to non-linear conversion factors given by
\begin{equation}
    \label{pertesNL}
    C_{i} = 1-\frac{\lambda_3 P_{3}}{\lambda_j P_j^{ \rm{cav}}},\quad (j\neq i;\ i,j=1,2).
\end{equation}

Assuming total transmission of the yellow light by the output mirror M4, the $589$~nm power $P_{3}$ coupled out of the cavity is given by $P_{3}\approx \alpha P_{1}^{\rm cav} P_{2}^{\rm cav}$, under the undepleted pump approximation. When both pumps have imbalanced powers, the output power at $\omega_{3}$ is ultimately limited by the weakest one, since one photon from both pumps is required to create one at $\omega_{3}$. As it is the case in our experiment, we assume that the weakest source is the one at wavelength $\lambda_2$. This translates into a maximum power $P_{3}^{\rm max}=(\lambda_{2}/\lambda_{3})P_{2}^{\rm inc}$. Therefore, a figure of merit to characterize the conversion efficiency is the ratio
\begin{equation}
    \label{eta}
   \eta= \frac{P_{3}}{P_3^{ \rm{max}}}=\frac{\lambda_2}{\lambda_3}\frac{P_{3}}{P_2^{\rm inc}}
\end{equation}
between the actual power and the absolute maximum power that can be obtained from the available pump power coupled into the cavity $P_{2}^{\rm inc}$. 

The problem at hand is thus to maximize $\eta$ for given cavity parameters $\delta_{i}, L_{i}, \alpha$. This amounts to balancing the input coupler reflectances $R_{1},R_{2}$ with the total loss per round trip,
including the non-linear conversion. This is usually termed
impedance matching~\cite{kaneda1997a}. In our case, finding the impedance matching
point is a coupled problem, since one should maximize simultaneously
both intensities in the cavity using (\ref{IcavNL}) and
(\ref{pertesNL}). This last equation is critically dependent on the
single-pass conversion coefficient $\alpha$. 

\subsection{Total conversion in an idealized lossless cavity}

Let us first study the case, where passive losses in the cavity and on the input coupler can be neglected ($\delta_{i},L_{i}=0$ in the above equations). The question to be answered is whether it is possible to convert all photons at $\lambda_{2}$ into photons at the harmonic at $\lambda_{3}$ (cavity conversion efficiency $\eta = 1$). In~\cite{mimoun2008a}, we showed that this is indeed the case for any value of the input coupler reflectance $R_{2}$. We recall here the argument for completeness. We look for a solution where the output flux at $\lambda_{3}$ and the incident flux at $\lambda_{2}$ are equal, $P_{3}/(\hbar\omega_{3})=P_{2}^{\rm inc}/(\hbar\omega_{2})$. According to Eq.(\ref{pertesNL}), this corresponds to $\mathcal{C}_{2} =1-\frac{P_{2}^{\rm inc}}{P_{2}^{\rm cav}}$. The cavity equation (\ref{IcavNL}) becomes
\begin{eqnarray}
P_{2}^{\rm cav} =\frac{(1-R_{2})P_{2}^{\rm inc}}{\left(1-\sqrt{R_{2}\left(1-\frac{P_{2}^{\rm inc}}{P_{2}^{\rm cav}}\right)}\right)^2}. 
\end{eqnarray}
This solves into the simple result
\begin{eqnarray}
P_{2}^{\rm cav} =\frac{P_{2}^{\rm inc}}{1-R_{2}},
\end{eqnarray}
valid for any $R_{2}$. Thus we conclude that there is always a possibility to reach complete conversion in the ideal, lossless case, corresponding to the intracavity flux for the weak pump $2$ as given above. The flux for the strong pump $1$ is found from the relation $P_{3}=\alpha P_{1}^{\rm cav} P_{2}^{\rm cav}$, 
\begin{eqnarray}\label{A1lossless}
P_{1}^{\rm cav} =\frac{(1-R_{2})\lambda_{2}}{\alpha_{\rm sp} \lambda_{3}}.
\end{eqnarray}
The parameters of the cavity ($R_{1},R_{2}$) are linked via Eq.(\ref{IcavNL}). For any $R_{2}$, one can find a value of $R_{1}$ leading to the power $P_{1}^{\rm cav}$ given above, corresponding to complete conversion of the $\lambda_2$ photons. 

\subsection{Optimization of conversion for a realistic lossy cavity}

In any practical situation, passive losses will be present. This modifies the conclusions of the last subsection, as these losses limit the enhancement factor that can be reached in the cavity. Unlike the lossless case, instead of a locus of optimal points in the ($R_1$,$R_2$) plane, one finds a unique value of ($R_1$,$R_2$) that maximizes $P_3$, at a value smaller than $P_3^{\rm max}$. However, this optimum is quite loose when non-linear conversion dominates over the passive losses (${\cal C}_i\gg\delta_i$). This highlights the importance of a large single-pass efficiency, justifying the use of a highly nonlinear material such as PPKTP : the required power $P_{1}$ is reduced (see Eq.(\ref{A1lossless})), making the cavity more tolerant to passive losses. 

We rewrite $\mathcal{C}_{2} =1-\eta \frac{P_{2}^{\rm inc}}{P_{2}^{\rm cav}}$. The cavity equation for wave $2$ then leads to two solutions for the intracavity power
\begin{eqnarray}\nonumber
P_{2}^{\rm cav} =\frac{(1+r)T_{2}-(1-r)r\eta}{(1-r)^2} \pm 2\frac{\sqrt{T_{2}r(T_{2}-(1-r)\eta)}}{(1-r)^2},
\end{eqnarray}
with $r=R_{2}(1-\delta_{2})$ the total passive loss coefficient for the circulating waves. The existence of two solutions indicates a possible bistability. Such solutions are real provided
\begin{eqnarray}
T_{2} \geq (1-r)\eta.
\end{eqnarray}
When this condition is not fulfilled, the cavity is unstable due to excessive passive or nonlinear losses. This condition sets a limit on the efficiency achievable for given cavity parameters $T_{2},R_{2},\delta_{2}$, $\eta < \eta^{\rm max}=T_{2}/(1-r)$. Assuming one chooses the input coupler to reach this maximum value, one finds for small losses ($\delta_{2}, L_{2}\ll R_{2}$) an intracavity power
\begin{eqnarray}
P_{2}^{\rm cav} \approx \frac{P_2^{\rm inc}}{1-R_{2}}\left(1-\frac{L_{2}+2R_{2}\delta_{2}}{1-R_{2}}\right),
\end{eqnarray}
 close to the idealized case studied before. 

\begin{table}[ht!]

	\caption{Reflectances and transmittances of the optical elements inside the cavity, at both wavelengths. Reflectances are measured within 0.5\% and transmittances within 0.2\%. The values for the crystal are specifications by the manufacturer. ${\cal R}$ is the optimum given by the numerical simulation.}

\begin{tabular}{r|ccc|ccc}
\multicolumn{7}{c}{}\\
\hline\hline
& \multicolumn{3}{c|}{~~1064~nm}& \multicolumn{3}{c}{~~1319~nm}\\
&R  & ${\cal R}$&T  & R & ${\cal R}$ & T   \\
\hline
$M_1$&  0.930 & 0.96&0.060 & 0.740&0.79 & 0.250  \\
$M_2, M_3, M_4$& 0.995& & 0.005& 0.995& &0.005\\
Crystal& &&0.980& &&0.980\\

\end{tabular}

\label{table:reflectivities}
\end{table}

\begin{figure*}[hptb]
    \centering
	\includegraphics{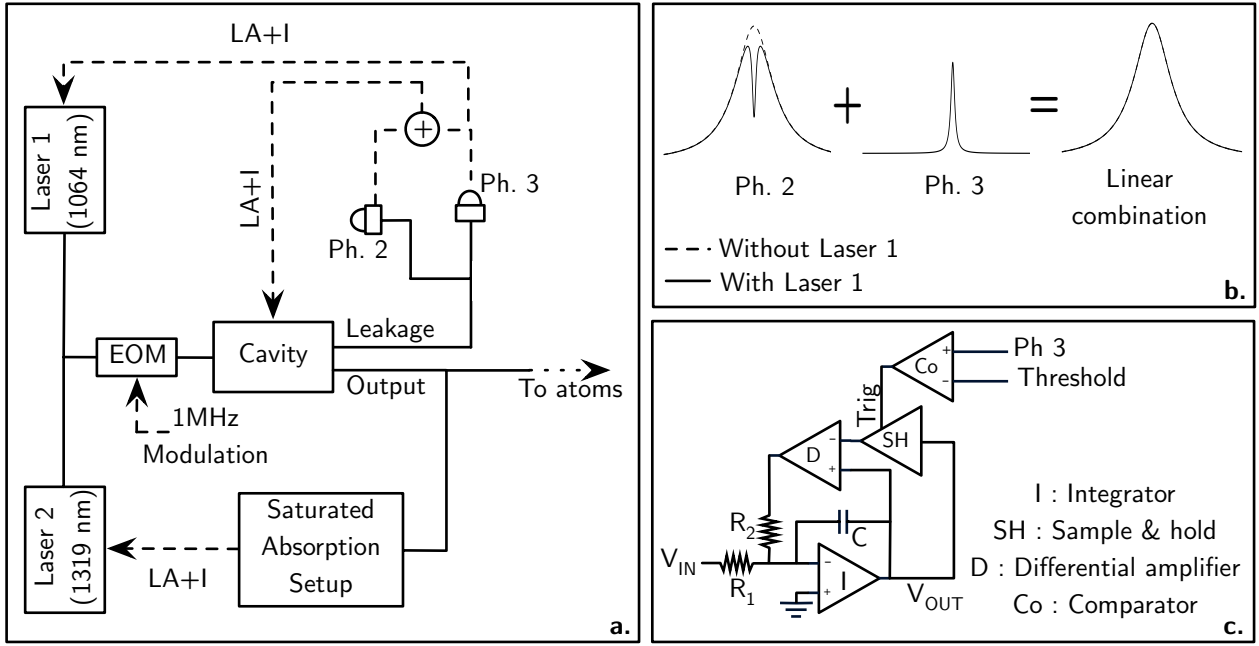}
    \caption {{\bf a}: Locking scheme for the SFG cavity. Solid lines represent optical paths, while dashed lines indicate electronic connections. LA + I: Use of a lock-in amplifier and an integrator to lock the lasers or the cavity to the maximum of a signal produced by one of the photodiode (Ph.). {\bf b}: Linear combination used to compensate for the dip in intra-cavity power due to conversion. The photodiode monitoring laser 2 sees a power drop while the photodiode monitoring laser 3 sees a peak. Summing them up allows one to always maintain a peak signal to lock to. {\bf c}: Automatic gain control circuit to control integrator saturation. COMP: comparator, SH: sample-and-hold amplifier, INT: integrator, D: precision differential amplifier.}
    \label{fig:schema_lock}
\end{figure*}

The solution of the coupled equations giving $P_{1}^{\rm cav},P_{2}^{\rm cav}$ [Eq.(\ref{IcavNL})] is performed numerically~\footnote{A numerical algorithm maximizing the two dimensional function $P_1^{\rm cav}P_2^{\rm cav}=f(R_1,R_2)$ was implemented. In practice, $P_1^{\rm cav}$ is first evaluated using an Euler secant method by substituting $P_2^{\rm cav}$ in $C_1$ [Eq.(\ref{pertesNL})] by its expression given by Eq.(\ref{IcavNL}). Once $P_1^{\rm cav}$ solved, its value is used to derive $P_2^{\rm cav}$ and the function $f$. The optimal couple ($R_1^{\rm opt}$,$R_2^{\rm opt}$) is then tracked using an adaptative stepsize algorithm maximizing $f$.}, using as input
the available power in our infrared sources and the measured characteristics of the cavity. We carried out the optimization with respect to the input coupler transmittances at both infrared wavelengths~\cite{mimoun2008a}.

Experimentally, we characterized carefully the transmition and reflection coefficients of the mirrors used for the cavity (see Table \ref{table:reflectivities}). The measured reflectances correspond to a passive ({\it i.e.} without non-linear conversion) amplification of the intra-cavity power by a factor around $22$ at $1064~$nm and $12$ at $1319~$nm. To find the powers coupled into the cavity fundamental mode, we sent the lasers independently into the cavity with a known incident power. Comparing the measured intracavity power (inferred from the power transmitted through M2 and the measured value of its transmittance) with the one expected from the reflectances gives the fraction of incident power effectively coupled to the fundamental mode, around $85\%$ for both wavelengths. The measured reflectances as well as coupling efficiencies were taken into account in our numerical simulations, predicting a conversion efficiency $\eta\simeq 0.9$ for the photons at $\lambda_2$ coupled into the cavity (see~\cite{mimoun2008a}). The maximal measured output power of 800~mW, which corresponds to $\eta=0.92$, is in fair agreement with this result.

\subsection{Cavity setup and locking}

An essential requirement to achieve a stable output with high efficiency is to ensure that both pump lasers are simultaneously resonant in the cavity.  In order to maintain the cavity on resonance for both wavelengths, a double locking scheme using the stable 1319~nm source as a master
laser is implemented~(see Fig.~\ref{fig:schema_lock}{\bf a}). Both lasers are routed together through an electro-optical modulator (EOM) placed before the cavity, and resonantly driven at a frequency $f_{\rm mod}=1~$MHz. This dithers the laser frequencies and generate a dispersive signal from the cavity transmission. In our implementation, the weaker pump laser $2$ is used as a
master laser onto which the cavity length is locked using an
integrating servo-loop. In a second step, the stronger pump $1$ laser is locked
onto the cavity, and consequently on the master laser, ensuring
stable operation of the ensemble. In details, the small fraction of
infrared light transmitted by the second mirror M$_2$ is collected
by two separate photodiodes (see Fig.~\ref{fig:Cavity}). Two
piezoelectric transducers glued to the cavity mirrors M$_2$ and
M$_3$ are used to tune the cavity length. The first one (M$_3$)
allows fast response in the 30~kHz range, but has a limited travel
of a few tens of nm. The second piezoelectric stack allows to
correct for larger drifts of the cavity length, occuring over much
longer timescales (from a few ms to a few hours). The photodiode signal at 1319~nm (Ph.2) is demodulated by a
lock-in amplifier operating at the modulation frequency $f_{\rm mod}$ driving the EOM, producing a dispersive error signal subsequently fed back to both
piezoelectric transducers (with appropriate filters in the feedback
loop). This locks the length of the cavity on the stable 1319~nm
source. The photodiode signal at 589~nm (Ph.3) is demodulated in a similar way, and the resulting signal can be used to react on laser 1 frequency using available piezoelectric and temperature control to ensure that it follows the cavity resonance. 

In situations where the conversion efficiency is large, this standard locking scheme leads to
serious stability problems with both IR lasers
simultaneously present in the cavity. To see this, picture a
situation where the cavity is on lock with the 1064 laser
off-resonant. As the 1064~nm laser frequency is tuned to reach
resonance, the power level of the 1319~nm drops due to conversion
into 589~nm photons (see Fig.~\ref{fig:schema_lock}{\bf b}). This large drop
of the 1319~nm power level when both lasers resonate cannot be
distinguished from a perturbation by the cavity lock. Hence, the cavity lock actually works against keeping both
lasers on resonance simultaneously, and resists increasing the
conversion efficiency above the level where the 1319~nm lineshape is
distorted significantly. We have devised a simple solution to this problem~\cite{mimoun2008a,brevet589}. First, instead of the bare $1319~$nm photodiode transmission, the error signal for the cavity lock is derived from a
linear combination of this transmission signal and of the yellow output of
the laser. The combination is done electronically before the lock-in
amplifier, with weights empirically chosen to minimize distortions of the cavity lineshape and to optimize the servo gain around the lock point. Our ``fringe reshaping'' method works for any level of conversion, and allows stable 
operation of the laser on a day time scale, even at the highest efficiencies. Second, choosing 
the $589~$nm output as the error signal for the second servo-loop instead of the $1064~$nm transmission ensures that the system locks to the maximal converted power. Our method relies on the fact that the SFG is a
phase-coherent process : a modulation sideband present on the
$1319~$nm laser is automatically present on the output (with a
different weight that depends on the $1064~$nm power). Synchronous demodulation by the lock-in amplifier
therefore preserves the linear combination. 

When these two servos are in action, the cavity is doubly resonant,
and the two lasers are frequency locked to each other. We have found
that the lock of the second laser was somewhat sensitive to
disturbances occurring near the optical table. This is a well-known
features of integrating servo-loops, which typically encounter
difficulties to recover from disturbances with large amplitude that
cause the integrator to saturate~\cite{fox2001a}. Integrators are required to achieve zero DC errors in a servo loop, and replacing them with a simpler proportional control is not an option. We have implemented an electronic circuit that bypasses this problem and prevents the integrator from saturating after violent perturbations, while maintaining the laser locked at all times. This can be seen as an automatic gain control circuit that limits the DC gain when the input becomes too large. The circuit, shown schematically in Fig.\ref{fig:schema_lock}{\bf c} , uses the laser power level to detect such disturbances, and compares it to a preset threshold value (set to 80\% of the nominal value in our case). A sample and hold amplifier (SH) samples the integrator voltage, with its output connected (``bootstrapped'') to the integrator input through a differential amplifier. Regular operations correspond to the SH in ``sample'' mode, where the output closely tracks the input. The output $V_{\rm diff}$ of the differential amplifier is zero and the integrator behaves normally. When the output of the laser falls below the threshold, the comparator triggers the SH circuit  to switch to ``hold''  mode. The SH output is frozen at the value $V_{\rm th}$ it had at threshold, so that the differential amplifier output becomes $V_{\rm diff}=V_{\rm out}-V_{\rm th}$. The output voltage at frequency $ \omega$ then becomes
\begin{equation}
V_{\rm out}=-\frac{\frac{R_{2}}{R_{1}}V_{\rm in}-V_{\rm th}}{1+i R_{2} C \omega},
\end{equation}
where $V_{\rm in}$ is the incoming error signal and $V_{\rm out}$ the output of circuit. The integrator is thus neutralized before saturating, and the circuit behaves as a proportional controller around the threshold value. When the perturbation is removed, the SH turns back to sample mode and restores regular integrator operation. We use this circuit on all servo controllers in the laser system.

With this last improvement, the system can withstand severe mechanical perturbations without unlocking and requires very little maintenance compared to dye lasers. It is mostly insensitive to temperature fluctuations because the infrared lasers are thermally stabilized and the cavity length fluctuations are compensated by the servo. Alignement is left untouched over weeks, with a power drop below 10\%, and when needed adjustments are only required on the injection path into the cavity. The cavity alignment itself has not been touched for six months. The laser stays locked for a day on its own, and for several hours when tracking an atomic line.

Finally, the frequency drift of the yellow laser is cancelled by locking laser 2 to the D$_2$ line of Na using standard saturated absorption spectroscopy, yielding a long-term frequency-stabilized laser source.

\section{Yellow laser characterization}
\label{sec:yellow}

\begin{figure}[htbp]
\includegraphics[width=8cm]{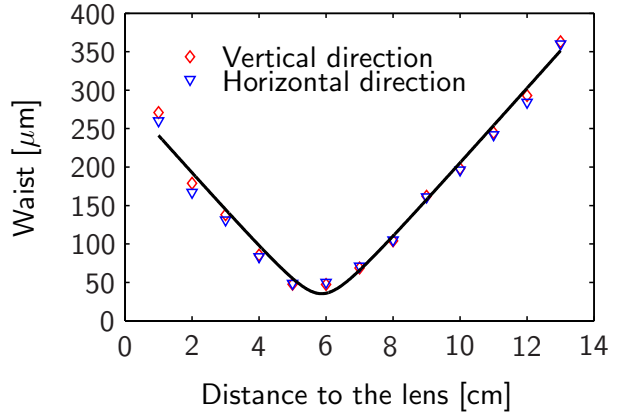}
\caption{Measurement of the M$^2$ coefficient for the 589~nm laser
in the vertical (triangles) and horizontal (circles) directions. The
laser is focused with a $f=100$~mm converging lens at the output of
the cavity.}
\label{fig:589nm_M2_measurement}
\end{figure}

\subsection{Beam quality}

We have characterized the spatial mode of the laser emerging from the cavity. The output beam was focused
through a converging lens and imaged on a charge-coupled device (CCD) camera at various distances from the lens. The beam profile for each distance was fitted to a gaussian with $1/e^2$ radius $w$ identified as the beam waist (see Fig.~\ref{fig:589nm_M2_measurement}). We fitted this
function to $w_{0}\sqrt{1+\theta (z/w_{0})^2}$, where $w_{0}$ is the
waist of the beam near focus, $\theta$ is its divergence, and $z$ the direction of
propagation. This gives a M$^2$ parameter M$^2=\pi w_{0} \theta/\lambda=1.02$, indicating diffraction-limited performances. This shows the high-quality of the transverse mode of the output beam.
Measurements in both transverse directions show no visible
astigmatism.

\subsection{Intensity noise measurements}

\begin{figure}[ht!]
\includegraphics[width=8cm]{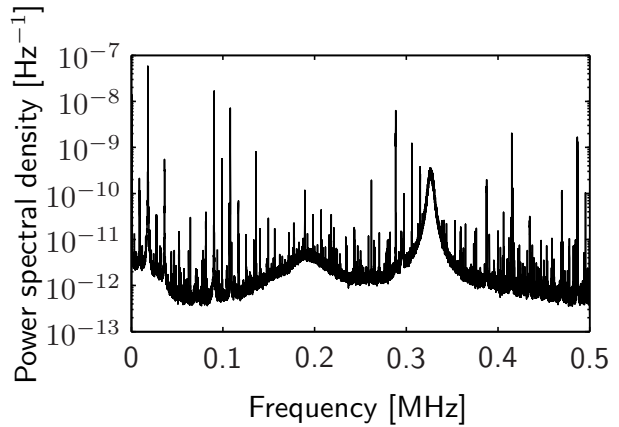}
\caption{Relative intensity noise spectral density of the laser source averaged over 100 samples.}
    \label{fig:psd}
\end{figure}

Intensity noise of the output was characterized by recording the beam power on a fast photodiode (bandwidth $10~$MHz) followed by a $16$ bits analog acquisition card (National Instruments NI-6259). From such samples, we determined the one-sided power spectral density $S_{\rm RIN}$ of the instantaneous intensity normalized to the mean intensity, 
\begin{equation}
S_{\rm RIN}(\nu) =\frac{1}{T} \left\langle\left \vert \int_{0}^T \frac{ I(\tau)}{\langle I \rangle} e^{i 2\pi \nu \tau} d\tau \right\vert^2\right\rangle,
\end{equation}
where $\langle\cdots \rangle $ denotes statistical averaging and where $T\approx 100$~ms is the measurement time. The results averaged over $100$ samples are shown in Fig.~\ref{fig:psd}. This corresponds to a noise $\delta I / \langle I \rangle\approx 4\times10^{-3}$ integrated over a $5$~Hz$-500$~kHz bandwidth. Two broad noise peaks are visible near $190$~kHz and $330$~kHz, which probably reflect resonances in the cavity piezoelectric actuators. The noise level is sufficient for our application, but could be controlled actively to a lower level if needed, for instance by monitoring the instantaneous power and reacting on the incident power from the $1064~$nm laser.

\subsection{Absorption from laser-cooled Sodium atoms}

\begin{figure}[ht!]
\includegraphics[width=8cm]{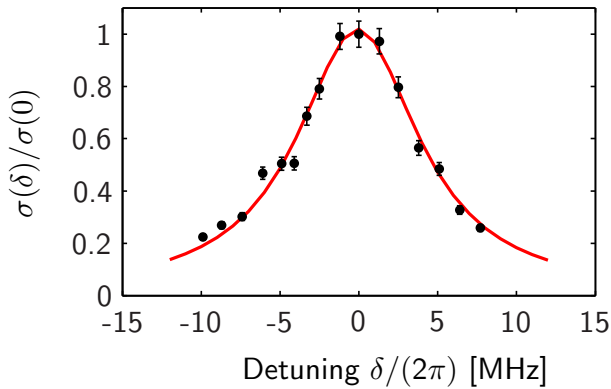}
\caption{Optical density of a cold atom cloud in a magneto-optical trap. The probe beam is detuned from the resonance line by an amount $\delta$.}
    \label{fig:scan_sonde}
\end{figure}

Because the two pump sources are extremely narrow in frequency, one can expect similar spectral purity of the output. At present times, we have no means to measure such narrow linewidths, but we can place an upper bound on the laser linewidth from high-resolution spectroscopy measurements. 

To this aim, we used cold atoms from a magneto-optical trap (MOT) formed using the SFG laser source. Sodium atoms were introduced in a high-vacuum cell using electrically-controlled dispenser sources (Alvatec GmbH). Repumping light was derived from the main laser using a high-frequency ($1.7~$GHz) acousto-optical modulator (Brimrose Corp.). A MOT was formed in the vacuum cell using a magnetic field gradient around $10~$G/cm and approximately $10$~mW optical power in each of the six MOT beams. The cloud typically contained a few $10^7$ atoms, at a temperature $T\sim 110~\mu$K. We measured the absorption of a weak probe beam (intensity $\sim 1~$mW/cm$^2$) by the atomic cloud (with MOT beams turned off) as a function of the probe frequency. According to Beer-Lambert's law, this measures $\sigma(\delta)$, the optical density at a detuning $\delta=\omega_{L}-\omega_{0}$, with $\omega_{L}$ the laser frequency and $\omega_{0}$ the atomic resonance frequency. Typical results are plotted in Fig.~\ref{fig:scan_sonde}. These measurements have been fitted using the theoretical expression  
\begin{equation}
\sigma(\delta)/\sigma(0)=\frac{\Gamma^2/4}{\delta^2+\Gamma^2/4},
\end{equation}
where $\Gamma$ is the natural linewidth for the $D_2$ transition of sodium. We deduce a measured value $\Gamma/(2\pi)\approx 9.6\pm0.5$~MHz, compatible with the value found in the literature, $\Gamma/(2\pi)=9.8$~MHz. Since no broadening of the absorption profile due to the linewidth of the laser could be observed within our measurement accuracy, the latter is small compared to the natural linewidth of the atoms. We conclude that the laser source fullfils all the requirements for laser cooling applications.

\section{Conclusion}

In conclusion, we have demonstrated a single-frequency, tunable, compact and robust all-solid-state SFG yellow laser source for cooling and trapping sodium atoms. The long-term stability of the laser source, despite the complexity brought by the use of a doubly-resonant enhancement cavity, stems from an original electronic servo loop. This servo is designed both to bypass the large depletion dip observed on the weaker input laser resonance fringe under high conversion, and to avoid saturation due to disturbances of the various integrators used in the servo loops. In the current configuration, the maximum output power of 800~mW remains lower than
what can be produced with a dye laser. However, based on our measurements,
we calculate that increasing the powers of the infrared laser sources to
$P_1=2$~W and $P_2=800$~mW (both commercially available) should allow
output powers in excess of $1$~W, ultimately limited by the possible occurrence of thermal effects in the crystal~\cite{targat2005a,torabi2003a,lundeman2008a}.

\begin{acknowledgement}
We would like to thank Pierre Lemonde, Wolfgang Ketterle and Aviv Keshet for
useful discussions. We acknowledge financial support from ANR
(contract Gascor), IFRAF (Microbec project), the European Union (MIDAS project, Marie Curie Fellowship) and DARPA (OLE project).
\end{acknowledgement}
\bibliographystyle{prstyNoEtAl}
\bibliography{somme_frequence}

%
\end{document}